# Inverse Design of Perfectly-Matched Metamaterials Via Circuit-Based Surrogate Models and the Adjoint Method

Shrey Thakkar, *Graduate Student Member*, *IEEE*, Jorge Ruiz-García, *Member*, *IEEE*, Luke Szymanski, *Member*, *IEEE*, Gurkan Gok, *Member*, *IEEE* and Anthony Grbic, *Fellow, IEEE*

*Abstract*—In this work, perfectly-matched metamaterials (PMMs) are described and combined with inverse design to realize broadband devices. PMMs are discretized metamaterials with anisotropic unit cells selected from a constrained design space, referred to as perfectly-matched media. PMMs exhibit the unique property that all their unit cells are impedance-matched to each other as well as to the host medium they are embedded within under all excitations. As a result, PMM devices rely on reflectionless refractive effects to achieve a prescribed function. This property enables true time delay performance and promises broadband capabilities. Two design examples are presented to demonstrate the potential of inverse-designed PMMs: a compact, broadband beam-collimator with a prescribed amplitude taper and a multi-input multi-output beamformer exhibiting zero scan loss.

*Index Terms*— Broadband, metamaterials, multiple input multiple output, inverse design, tensor materials, adjoint optimization

## I. INTRODUCTION

THE metamaterial paradigm has grown to incorporate unprecedented degrees of freedom, enabling compact electromagnetic (EM) devices that can tailor EM responses at a subwavelength scale [1], [2], [3], [4]. Concurrent advances in fabrication techniques, in turn, have enabled their realization. Consequently, there has been rapid progress in the design of metamaterials and metasurfaces that perform elaborate field transformations [5], [6], [7], [8], [9] or facilitate multi-input multi-output (MIMO) capabilities [10], [11], [12]. However, analytical methods to achieving increasingly complex functionality can be limiting for several reasons. They typically rely on simplifications and are unable to incorporate practical design constraints, which often leads to underutilizing the dimensionality of the design space.

Alternatively, a device function can be posed as the solution to an inverse design problem [12], [13], [14]. This involves formulating the desired electromagnetic response as an objective function with a scalar figure of merit. Optimization-based inverse design methods aim to progressively minimize such objective functions. Specifically, they parse through a high-dimensional design space to identify non-intuitive solutions while subject to user-defined constraints such as practical considerations. Previously, metamaterials have been realized using this technique for a wide range of applications, such as analog signal processors, beamformers, high-frequency circuit components, and optical technology, including optical neural networks [11], [12], [13], [14], [15], [16], [17], [18], [19], [20], [21], [22].

The inverse design of these devices entails repeatedly computing the electromagnetic response of an inhomogeneous, multiwavelength structure possessing subwavelength scatterers. This is a non-trivial numerical problem that imposes practical constraints on the inverse design routine. The design procedures in [18], [20] employ full-wave solvers to accurately evaluate the structure's response in each iteration. Nevertheless, the associated high computational cost restricts their ability to engineer large structures with intricate geometries. Alternatively, deep-learning-based strategies have been reported [17], where a convolutional neural network (CNN) rapidly predicts the scattering properties of an arbitrary pixelated geometry. However, this approach transfers the computational cost to training the CNN, where extensive data sets of accurate EM simulations are required. Instead, a computationally effective strategy was proposed in [11] that avoids full-wave simulations in every iteration. The metamaterial was approximated as a circuit network, and its EM response rapidly evaluated using circuit theory. Circuit networks can capture crucial phenomena such as intercell coupling and spatial dispersion that are typically observed in inhomogeneous, anisotropic structures [11], [23], [24], [25]. For these reasons, a fast 2-D circuit network solver (CNS) has been adopted in this work.

Notably, inverse-designed devices achieve the desired response primarily influenced by the library of available constitutive elements/cells and their interaction with each other. For instance, a device may exploit inter-cell reflections to achieve a field profile with prescribed amplitude and phase [11]. This can lead to internal standing waves that create localized field hot spots, contributing to loss and narrow operating bandwidths. Here, this issue is addressed by

This work was supported by AFOSR Grant FA9550-24-1-0098. (Corresponding author: Anthony Grbic.)

S. Thakkar, J. Ruiz-García, and Anthony Grbic are with the Radiation Laboratory, Department of Electrical Engineering and Computer Science, University of Michigan, Ann Arbor, MI 48109 USA. (e-mail: shreyt@umich.edu; jruizgar@umich.edu; agrbic@umich.edu).

Luke Szymanski is currently an MIT Lincoln Laboratory, Lexington, USA employee. No Laboratory funding or resources were used to produce the result/findings reported in this publication. (e-mail: luke.szymanski@ll.mit.edu).

Gurkan Gok is with RTX Technology Research Center, East Hartford, CT 06108 USA. (e-mail: gurkan.gok@rtx.com).

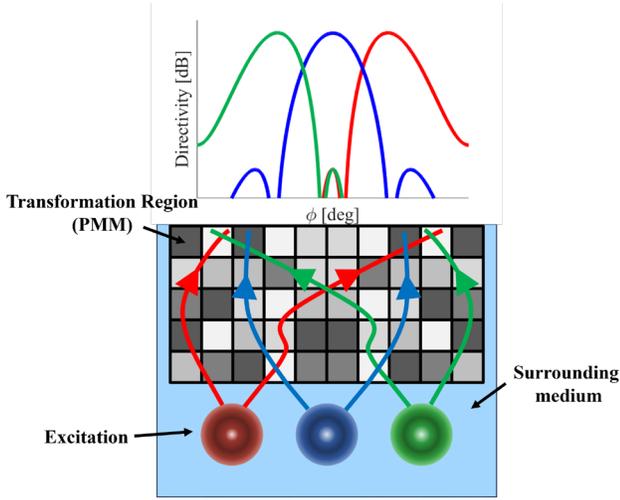

**Fig. 1.** A 2-D perfectly-matched metamaterial that performs true time delay multi-input multi-output field transformations.

constraining the material parameter space to a passive, lossless, and anisotropic set of media referred to as "perfectly-matched media" [25], [26], [27], [28], [29]. They possess the unique property of enabling all-angle reflectionless transmission across an interface separating any two media within this family [26]. Homogeneous blocks of perfectly-matched media placed in a 2-D lattice as shown in Fig. 1, are referred to as a perfectly-matched metamaterial (PMM). PMMs are inhomogeneous, anisotropic structures ($\varepsilon_z(x,y)$ and $\bar{\bar{\mu}}(x,y)$) that manipulate internal path lengths via refractive effects to perform a prescribed electromagnetic function.

PMMs are intimately related to transformation optics (TO) [30], [31], [32], [33], [34]. TO involves transforming an initial field distribution to a desired one through a coordinate transformation that distorts physical space. The distorted space can be realized through changes in the permittivity and permeability of the underlying medium. The unique property of TO devices is that they are also devoid of internal reflections and generally remain impedance-matched to the host (surrounding) medium. However, TO devices can be difficult to fabricate due to the continuous material parameter profiles that may also possess extreme values [34], [35], [36]. Additionally, the only methods to designing MIMO TO devices rely on physical symmetry, which severely limits their functionality. Analytical methods to designing PMMs [28] and TO-based metamaterials [37] that are uniformly discretized have been reported, but are limited in capability to single-input single-output (SISO) responses. Further, [18] reports an FEM-enabled inverse-design procedure that optimizes a uniformly discretized transformation region to design broadband MIMO devices. However, a fundamental constraint in the optimization routine restricts the devices to slowly varying material parameters. This limits the design space and functional scope, leading to electrically-larger devices.

This work improves on the procedure reported in [18] on two key fronts. Instead of a full-wave solver, the optimization routine is integrated with a CNS to enable rapid forward simulations. Further, confining the design space to perfectly-matched media mitigates internal reflections, allowing the inhomogeneity within the structure to be unrestricted. A gradient-based optimization routine is employed to design PMMs. A circuit representation for unit cells of PMMs has been initially reported in [27], [29] using tensor transmission-line (TLIN) metamaterials [38]. Tensor TLIN metamaterials allow material parameter distributions of $\epsilon_z(x,y)$ and $\bar{\bar{\mu}}(x,y)$ to be translated to electrical networks. As a result, the PMM's response can be evaluated using a custom in-house CNS, based on the one reported in [11]. Lossy, uniaxial variants of tensor TLIN unit cells are used to impose an absorbing boundary condition (ABC) at the edges of the computational domain within the CNS, similar to perfectly-matched layers [39], [40]. It is observed that the CNS solutions closely follow full-wave simulations of equivalent material parameter distributions: $\epsilon_z(x,y)$ and $\bar{\bar{\mu}}(x,y)$. This allows the inverse-design scheme to rapidly and accurately solve the forward problem in every iteration. The gradient-based optimization is further accelerated by the adjoint variable method [41], [42].

The procedure presented here has key advantages over the TO techniques mentioned earlier. First, it enables the design of devices whose coordinate transformations would be difficult to find, if a solution even exists. These include MIMO devices where independent and precise control of output fields for each input is demanded. Second, constraints can be imposed on the material parameters to account for practical considerations or improve bandwidth. Specifically, constraining the design space to perfectly-matched media ensures that the PMM is impedance-matched internally and to the host medium, regardless of the inhomogeneity. This permits the use of the entire design space to find non-intuitive solutions to complex field transformations that support broadband performance. This is unlike the procedures reported in [11], [18], where the device either suffers from narrowband performance or is restricted to spatially adiabatic material parameter distributions. Methods to realize the resultant material parameter distributions have been reported [18], [24], [25]. These devices can find applications where wideband, inverse-designed devices that rely on refractive effects are desirable, such as broadband signal routing, lensing, beamforming, and analog signal processing.

This paper is arranged as follows. Section II reviews the all-angle reflectionless transmission property of perfectly-matched media and presents the general form of their material parameters. Next, an ideal 2D TLIN unit cell that models perfectly-matched media is introduced. Section III discusses the inverse-design procedure employed in this work. Finally, the utility of PMMs is demonstrated through two design examples: 1) a compact, collimator that produces a beam with a cosine amplitude taper, and 2) a 3-input 3-output beamformer that azimuthally scans a cosine-tapered beam with no scan loss. Concluding remarks follow.

## II. PERFECTLY-MATCHED METAMATERIALS

PMMs are inhomogeneous devices composed of anisotropic unit cells selected from a family of perfectly-matched media. First, this section derives the material parameters of a perfectly-matched medium from its all-angle impedance matching property. Next, a tensor TLIN unit cell that models a homogeneous block of perfectly-matched media under S-polarized TEM excitation is reported for application in the CNS

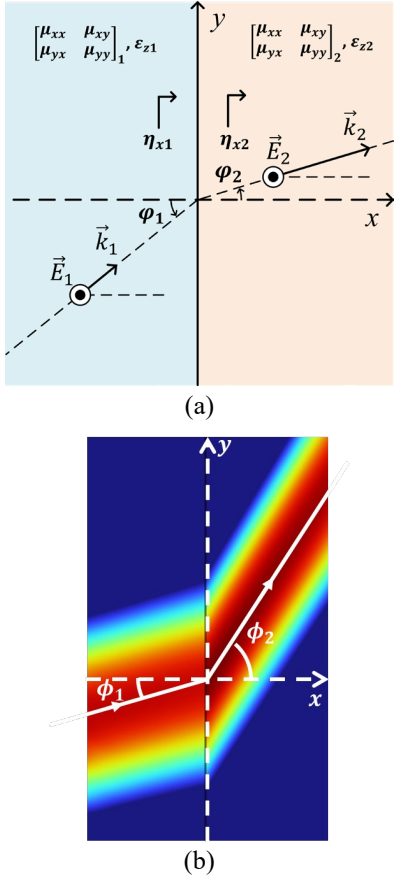

**Fig. 2.** Reflectionless transmission across a planar interface separating two homogeneous, lossless, magnetically anisotropic perfectly-matched media. No variation is assumed along the $z$ axis. a) Graphical illustration of the media, b) Example showing the transmission of a Gaussian beam ($|E_z|^2$).

*A. Review of Perfectly-Matched Media*

This section derives the properties of perfectly-matched media. First, let us consider an interface between two anisotropic half spaces and examine the conditions of perfect transmission for any incident angle. Consider an S-polarized TEM plane wave incident on a planar interface ($y$-$z$ plane) separating the homogeneous half-spaces of media 1 and 2, as shown in Fig. 2(a). Both media are assumed to be lossless, reciprocal, and magnetically anisotropic. The plane wave has nonzero field quantities $E_z$, $H_x$ and $H_y$, and it is incident at an arbitrary angle $\phi_1$ with respect to the $x$-axis. The material parameters relevant to the polarization of interest are a 2×2 relative permeability tensor in the $x$-$y$ plane and a relative scalar permittivity along the $z$ axis given by

$$\bar{\bar{\mu}}_i = \begin{bmatrix} \mu_{xx} & \mu_{xy} \\ \mu_{yx} & \mu_{yy} \end{bmatrix}_i, \quad \varepsilon_i = \varepsilon_{zi} \quad (1)$$

where $i = 1$ or $2$ for media 1 and 2, respectively. Next, the conditions for reflectionless transmission across the interface are examined.

Snell's law stipulates that the tangential wave number $k_y$ is continuous across the interface. That is, $(k_y)_1 = (k_y)_2 = k_y$, where $(k_y)_i = (k_i)\sin(\phi_i)$ and $\phi_i$ is the angle between $k_i$ and $\hat{x}$. Reflectionless transmission of the plane wave occurs when the normal wave impedance $\eta_x = -\frac{E_z}{H_y}$, seen by the incident and refracted wave, are identical in both media, $(\eta_x)_1 = (\eta_x)_2$. Therefore, to ensure reflectionless transmission for all incidence angles (perfect matching), the following condition must hold

$$(\eta_x)_1 = (\eta_x)_2 \quad \forall \quad k_y. \quad (2)$$

The plane wave relationship between $k_y$ and $\eta_x$ can be derived from Faraday's and Ampere's laws for time-harmonic fields,

$$-j\begin{bmatrix} k_x \\ k_y \end{bmatrix} \times E_z\hat{z} = -j\omega\bar{\bar{\mu}}\mu_o \begin{bmatrix} H_x \\ H_y \end{bmatrix}, \quad (3a)$$

$$-j\begin{bmatrix} k_x \\ k_y \end{bmatrix} \times \begin{bmatrix} H_x \\ H_y \end{bmatrix} = j\omega\varepsilon_z\varepsilon_o E_z\hat{z}. \quad (3b)$$

These equations can be rewritten in terms of the normalized wave vectors and wave impedances [37],

$$\begin{bmatrix} \bar{k}_x \\ \bar{k}_y \end{bmatrix} = |\bar{\bar{\mu}}| \begin{bmatrix} \mu_{xx} & \mu_{xy} \\ \mu_{yx} & \mu_{yy} \end{bmatrix}^{-1} \begin{bmatrix} \bar{\eta}_x^{-1} \\ \bar{\eta}_y^{-1} \end{bmatrix}, \quad (4a)$$

$$\varepsilon_z = \begin{bmatrix} \bar{k}_x & \bar{k}_y \end{bmatrix} \begin{bmatrix} \bar{\eta}_x^{-1} \\ \bar{\eta}_y^{-1} \end{bmatrix}, \quad (4b)$$

where $|\bar{\bar{\mu}}|$ is the determinant of the relative permeability tensor, $\bar{\eta}_x = -\frac{1}{\eta_o}\frac{E_z}{H_y}$, $\bar{\eta}_y = \frac{1}{\eta_o}\frac{E_z}{H_x}$, $\bar{k}_x = \frac{k_x}{k_o}$, and $\bar{k}_y = \frac{k_y}{k_o}$ ($\eta_o$ and $k_o$ are the free space wave impedance and wavenumber respectively). The dispersion relation can be determined from (4a) and (4b) as

$$\bar{k}_x^2 \frac{\mu_{xx}}{|\bar{\bar{\mu}}|} + 2\bar{k}_x\bar{k}_y \frac{\mu_{xy}}{|\bar{\bar{\mu}}|} + \bar{k}_y^2 \frac{\mu_{yy}}{|\bar{\bar{\mu}}|} = \varepsilon_z. \quad (5)$$

Note that (4a) assumes that the permeability tensor is invertible and (5) assumes a reciprocal medium ($\mu_{xy} = \mu_{yx}$). Using (4a), the normal wave impedance can be written as

$$\bar{\eta}_x^{-1} = \bar{k}_x \frac{\mu_{xx}}{|\bar{\bar{\mu}}|} + \bar{k}_y \frac{\mu_{xy}}{|\bar{\bar{\mu}}|}, \quad (6a)$$

$$\bar{\eta}_y^{-1} = \bar{k}_y \frac{\mu_{yy}}{|\bar{\bar{\mu}}|} + \bar{k}_x \frac{\mu_{xy}}{|\bar{\bar{\mu}}|}. \quad (6b)$$

Finally, using (5) and (6a), $\bar{k}_y$ can be related to $\bar{\eta}_x$ as follows,

$$\frac{\left(\frac{1}{\bar{\eta}_x}\right)^2}{\left(\sqrt{\frac{\mu_{xx}\varepsilon_z}{|\bar{\bar{\mu}}|}}\right)^2} + \frac{(\bar{k}_y)^2}{(\sqrt{\mu_{xx}\varepsilon_z})^2} = 1. \quad (7)$$

Perfect matching can be enforced between the two media using expressions (2) and (7), which results in the following two conditions [26],

$$|\bar{\bar{\mu}}|_1 = |\bar{\bar{\mu}}|_2 = \Delta, \quad (\mu_{xx}\varepsilon_z)_1 = (\mu_{xx}\varepsilon_z)_2 = \kappa, \quad (8)$$

where $|\bar{\bar{\mu}}|_1$ and $|\bar{\bar{\mu}}|_2$ are the determinants of the permeability tensors in media 1 and 2 respectively, and $(\mu_{xx}\varepsilon_z)_1$ and $(\mu_{xx}\varepsilon_z)_2$ are the normal refractive indices in each medium. Applying the conditions imposed by (8) to (1), the relative material parameters of perfectly-matched media can be cast in

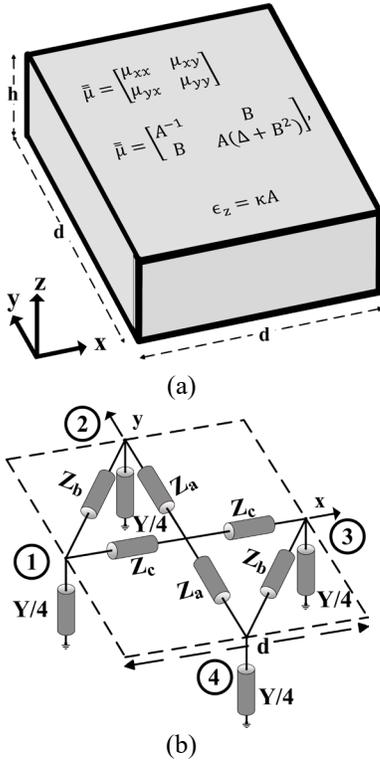

**Fig. 3.** Unit cell models for perfectly-matched media: a) unit cell that is a homogeneous, anisotropic block of material, b) equivalent tensor TLIN unit cell of a bowtie topology

the following general form,

$$\bar{\bar{\mu}} = \begin{bmatrix} A^{-1} & B \\ B & A(\Delta + B^2) \end{bmatrix}, \quad \varepsilon_z = \kappa A, \quad (9)$$

where $A$ and $B$ are free variables, and constants $\Delta$ and $\kappa$ are determined by the host medium. Setting $A$ and $B$ to real values results in lossless and passive perfectly-matched media. How blocks of such media can be combined to constitute PMMs is explained in the next paragraph. In contrast, setting $A$ and $B$ to complex values results in lossy and non-passive variants of perfectly-matched media. Their insensitivity to incidence angle combined with the presence of loss makes them suitable for terminating computational domains [29], much like PMLs [43]. A thorough investigation of this strategy is deferred to a future publication.

In PMMs, each unit cell is a discrete block of homogeneous, perfectly-matched media with electromagnetic properties defined by $A$, $B$, $\Delta$ and $\kappa$, as given in (9) and illustrated in Fig. 3(a). Imposing a common $\Delta$ and $\kappa$ for all discrete blocks ensures that the PMM is impedance-matched internally. When $\Delta$ and $\kappa$ are properties shared by a host isotropic medium, the PMM is also perfectly-matched to the surroundings for all illuminations [36].

In summary, homogeneous blocks having material parameters of the form described by (9) are perfectly-matched: impedance matched to each other across a $y$-directed interface, as well as to the host medium for all S-polarized TEM excitations. The general form of the material parameters for media that are perfectly-matched across $x$-directed interfaces can be derived in a similar manner. Their relative permeability tensor is given by (9) and the relative permittivity is

$$\varepsilon_z = \frac{\kappa}{A(\Delta + B^2)}. \quad (10)$$

The following subsection derives a circuit representation of the PMMs' unit cells.

*B. Transmission-Line Model of Perfectly-Matched Media*

The proposed inverse design procedure employs reduced-order models (four-port admittance matrices) of the metamaterial's unit cells. It computes the forward problem by modeling the unit cell interactions using circuit theory. In 2-D configurations, each unit cell can be considered a block of perfectly-matched media within a parallel-plate waveguide under S-polarized TEM excitation [25]. Tensor TLIN unit cells have been previously used to model such arbitrary tensor materials supporting S-polarized TEM fields [38]. In this section, we explicitly derive the parameters of a tensor TLIN unit cell that models perfectly-matched media.

A square, homogeneous, and anisotropic block with material parameters given by (9) is depicted in Fig. 3(a). Also, a tensor TLIN unit cell (shunt node configuration) is shown in Fig. 3(b). The series impedances $Z_c$, $Z_a$, and $Z_b$ are arranged in a bowtie topology. They lie along the $x$ and $y$ axes, and $x$-$y$ diagonal, respectively. The scalar shunt admittance $Y$ represents the shunt branch of the network. In the long wavelength regime ($\lambda_0 \gg d$, where $\lambda_0$ is the free-space wavelength and $d$ the unit cell's dimension), the TLIN unit cell shown in Fig. 3(b) can be characterized in terms of a series impedance tensor and scalar admittance,

$$\bar{\bar{Z}} = \begin{bmatrix} Z_{xx} & Z_{xy} \\ Z_{yx} & Z_{yy} \end{bmatrix}, \quad Y. \quad (11)$$

For the polarization of interest, we can establish a relationship between anisotropic, homogeneous media and a periodic network composed of the TLIN unit cell in Fig. 3(b) [38],

$$j\omega d \begin{bmatrix} \mu_{yy} & -\mu_{xy} \\ -\mu_{yx} & \mu_{xx} \end{bmatrix} = \begin{bmatrix} Z_{xx} & Z_{xy} \\ Z_{yx} & Z_{yy} \end{bmatrix}, \quad (12a)$$

$$j\omega d \varepsilon_z = Y. \quad (12b)$$

As mentioned above, $d$ represents the periodicity of the uniformly discretized PMM and its equivalent circuit network. When the phase delay across the unit cell is small (deep subwavelength limit, $k_x d, k_y d \ll 1$), the impedance tensor $\bar{\bar{Z}}$ can be expressed in terms of the series lumped impedances,

$$\bar{\bar{Z}} = \begin{bmatrix} \dfrac{2Z_c(Z_a + Z_b)}{Z_a + Z_b + Z_c} & -\dfrac{2Z_a Z_c}{Z_a + Z_b + Z_c} \\ -\dfrac{2Z_a Z_c}{Z_a + Z_b + Z_c} & \dfrac{2Z_a(Z_c + Z_b)}{Z_a + Z_b + Z_c} \end{bmatrix}. \quad (13)$$

The series impedances of the equivalent tensor TLIN unit cell can be written in terms of the variables $A$, $B$ and $\Delta$ using (9) and (13) in (12a), leading to

$$Z_a = \frac{j\omega d \Delta}{2[A(\Delta + B^2) - B]}, \quad (14a)$$

$$Z_b = \frac{j\omega d \Delta}{2B}, \tag{14b}$$

$$Z_c = \frac{j\omega d A \Delta}{2(1-AB)}. \tag{14c}$$

The shunt admittance can be derived from (9) and (12b) to be

$$Y = j\omega d \kappa A. \tag{14d}$$

Setting $A$ and $B$ to real values results in imaginary series and shunt lumped impedances. This leads to circuit networks that are lossless, reciprocal, and passive. Alternatively, complex-valued $A$ and $B$ lead to complex shunt and series impedances, resulting in lossy or nonpassive networks.

The 4×4 admittance matrix entries of the tensor TLIN unit cell in Fig. 3(b) can be written in terms of the unit cell impedances as,

$$Y_{11} = Y_{33} = \frac{1}{4}\left(Y + 2\left(\frac{1}{Z_c} + \frac{2}{Z_b} + \frac{1}{Z_a + Z_c}\right)\right), \tag{15a}$$

$$Y_{22} = Y_{44} = \frac{1}{4}\left(Y + 2\left(\frac{1}{Z_a} + \frac{2}{Z_b} + \frac{1}{Z_a + Z_c}\right)\right), \tag{15b}$$

$$Y_{21} = Y_{34} = -\frac{1}{2(Z_a + Z_c)}, \tag{15c}$$

$$Y_{31} = -\frac{Z_a}{2Z_c(Z_a + Z_c)}, \tag{15d}$$

$$Y_{41} = Y_{32} = -\frac{1}{Z_b} - \frac{1}{2(Z_a + Z_c)}, \tag{15e}$$

$$Y_{42} = -\frac{Z_c}{2Z_a(Z_a + Z_c)}, \tag{15f}$$

where the indices refer to the unit cell ports (nodes) at the center of each unit cell edge (see Fig. 3(b)). The remaining terms are determined from (15) via reciprocity. Hence, a homogeneous block of any magnetically anisotropic media with properties given by (1) can be represented as the admittance matrix of this TLIN unit cell using (12), (13), and (15). Specifically, the equivalent admittance matrix of a perfectly-matched medium defined by (9) can be derived via (14) and (15) as a function of the medium's variables $A$, $B$, $\Delta$, and $\kappa$.

Similarly, TLIN unit cells representing media that are perfectly-matched across $x$-directed interfaces possess series impedances given by (14a-c) and a shunt admittance given by,

$$Y = \frac{j\omega d \kappa}{A(\Delta + B^2)}. \tag{16}$$

III. INVERSE DESIGN PROCEDURE

The inverse design method applied in this work consists of two modules: the forward solver and the optimization routine. First, the fast 2-D CNS used to analyze the metamaterial (the forward problem) is presented. In addition, the transmission-line-based ABCs used to terminate the computational domain are introduced. The gradient-based optimization routine, which iteratively updates the design variables, is subsequently outlined.

A. 2-D Circuit Network Solver

In this section, a frequency-domain solver for 2-D PMMs supporting S-polarized TEM fields is presented. The inverse design routine requires solving the forward problem in every iteration of the optimization. Indeed, full-wave methods are often used to simulate a structure's response accurately [18], [20], [44]. Alternatively, deep-learning-based models have been used to predict scattering parameters rapidly. However, their training requires extensive datasets of accurate EM simulations [17]. The computational cost associated with both techniques scales with the size and geometry of the structure, as well as the span of the design space. To circumvent these constraints, surrogate models that approximate full-wave solutions have been developed [11], [15]. Particularly in [11], the metamaterial is approximated as an electrical network constructed by interconnecting reduced-order four-port admittance matrices that approximate the metamaterial's unit cells. The admittance matrices are extracted via full-wave simulations of isolated, realizable transmission-line unit cell geometries. Subsequently, to solve the forward problem, a CNS applies Kirchoff's Current Law (KCL) at every node in the network. This results in a linear system whose solution consists of nodal voltages. For electrically-large structures, the linear system is sparse, allowing for the rapid computation of an aperiodic metamaterial's response

In this work, the 2-D CNS presented in [11] is extended to analyze uniformly discretized transformation regions (PMMs) possessing inhomogeneous and magnetically anisotropic material parameter distributions ($\varepsilon_z(x,y)$ and $\bar{\bar{\mu}}(x,y)$) under S-polarized TEM excitation. Specifically, the computational domain (depicted in Fig. 4(a)) consists of an M×N lattice of unit cells with period $d$. The PMM (grayscale blocks) is embedded within free space (blue blocks), which is, in turn, enclosed by transmission-line-based ABCs (green blocks) that form the boundaries of the computational domain. Each unit cell is represented by a 4×4 admittance matrix, given by (15), as described in Section II-B. The construction of these ABCs using TLINs is discussed in the following section. Thus, the entire domain illustrated in Fig. 4(a) is modeled as a network composed of tensor TLIN unit cells interconnected at the unit cell ports. The resultant TLIN network is shown in Fig. 4(b). Excitations are imposed at some specific unit cell ports using a lumped voltage source in series with a lumped impedance. Consequently, the inverse design routine optimizes 2-D electrical networks (enclosed by the red dashed lines in Fig 4(b)). The optimized networks are ultimately translated into effective material parameter distributions ($\varepsilon_z(x,y)$ and $\bar{\bar{\mu}}(x,y)$), represented by the grayscale blocks in Fig. 4(a). At this stage, a database can be developed to relate the required material parameters to realizable geometries. References [23], [25] outline two approaches to realizing these optimized material parameter distributions.

B. Transmission-Line-Based Absorbing Boundary Conditions

Aperiodic structures support a spectrum of plane waves, which are incident at various angles on the boundary of the computational domain. If the boundaries are terminated with a fixed impedance, as in [11], [27], these obliquely incident waves undergo partial reflections and are reintroduced into the computational domain. This leads to inaccuracies in the analysis of the isolated metamaterial. In this work, the absorbing boundary layers are lossy, uniaxial, TLIN-based, and implemented in a fashion similar to [40].

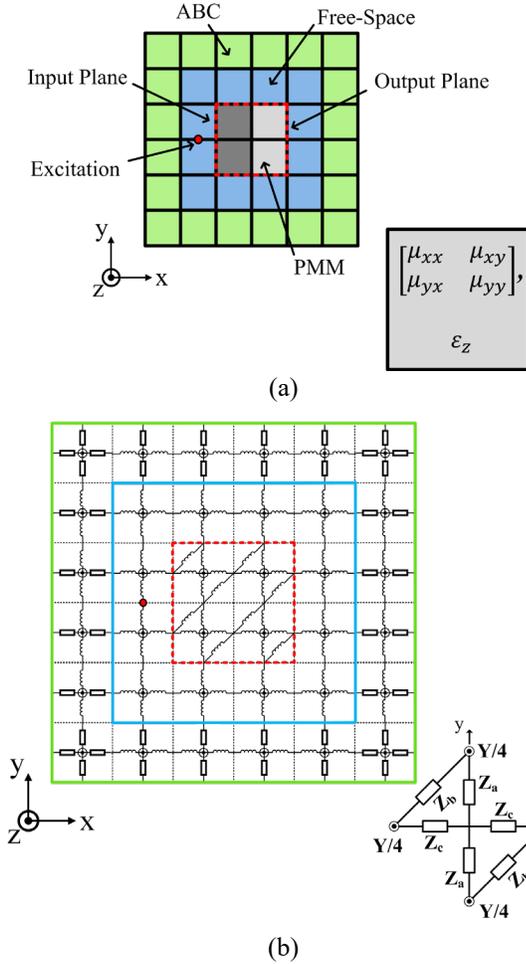

**Fig. 4.** 2-D computational domains for analyzing a PMM using a a) full-wave solver and b) Circuit Network Solver (CNS).

Let us consider the interfaces parallel to $\hat{y}$ in Fig. 4, that separate the free space region (blue), from the ABCs (green). A desirable ABC is one that introduces loss in the normal direction ($\hat{x}$), while minimizing the normal wave impedance mismatch with free space for all angles of incidence. Let us assume the following relative material parameters for the uniaxial ABCs under consideration,

$$\bar{\bar{\mu}}_{y-abc} = \begin{bmatrix} 1 & 0 \\ 0 & 1-j\delta \end{bmatrix}, \quad \varepsilon_{z-abc} = 1-j\delta, \quad (17)$$

where $\delta$ is a real and positive value that represents a loss factor. Substituting $k_x$ from (5) into (6), and imposing $\mu_{xy} = 0$, the normal wave impedance, $\eta_x$, for the medium becomes

$$\eta_x = \frac{\omega\mu_0}{\sqrt{k_0^2 \frac{\varepsilon_z}{\mu_{xx}} - k_y^2 \frac{1}{\mu_{xx}\mu_{yy}}}}. \quad (18)$$

Then, substituting the material parameters given by (17) into (18) results in,

$$\eta_{x-abc} = \frac{\omega\mu_0}{\sqrt{k_0^2 - k_y^2 - j\delta k_y^2}}. \quad (19)$$

Note that the mismatch in the normal wave impedance is captured by the $j\delta k_y^2$ term, which increases with the incident angle. The loss term $\delta$ must be judiciously selected to balance the attenuation of the wave against the risk of significant impedance mismatch. Similarly, the absorbing layers parallel to $\hat{x}$ take on the following relative material parameters,

$$\bar{\bar{\mu}}_{x-abc} = \begin{bmatrix} 1-j\delta & 0 \\ 0 & 1 \end{bmatrix}, \quad \varepsilon_{z-abc} = 1-j\delta. \quad (20)$$

Finally, the corner absorbing regions, that exhibit loss along $\hat{x}$ and $\hat{y}$, are chosen to have the following relative material parameters,

$$\bar{\bar{\mu}}_{xy-abc} = \begin{bmatrix} 1-j\delta & 0 \\ 0 & 1-j\delta \end{bmatrix}, \quad \varepsilon_{z-abc} = 1-j\delta. \quad (21)$$

The absorbing regions are mapped to the bowtie tensor TLIN unit cell shown in Fig. 3(b) using (12), (13) and (15) and depicted in Fig. 4(b). Their effectiveness is crucial to the reliability of the inverse design procedure. It has been verified by comparing forward solutions computed using the 2-D CNS (Fig. 4(b)) and COMSOL Multiphysics [45]. Note that COMSOL simulates the scenario shown in Fig. 4(a), which is formed from homogeneous blocks and excited by line sources.

*C. Objective Function*

A MIMO device's functionality is described by a set of input-output pairs. In the context of this paper, input refers to the lumped voltage source excitations and output refers to all nodal voltages of the circuit network within the computational domain. The $n$th input-output pair can be written as $(\mathbf{v}_{in}^n, \mathbf{v}_{out}^n)$, where $n \in \{1, 2, \ldots, N\}$ and $N$ is the total number of pairs. To realize a desired functionality, a cost function is required to represent the device's performance for each input-output pair. The cost function is defined as follows [11],

$$g_n(\mathbf{p}) = \frac{1}{2}(\mathbf{v}^n(\mathbf{p}) - \mathbf{v}_{out}^n)^H \bar{\bar{G}}(\mathbf{v}^n(\mathbf{p}) - \mathbf{v}_{out}^n), \quad (22)$$

where $\mathbf{p}$ is a vector containing all the design variables $(A, B)$ and $\mathbf{v}^n(\mathbf{p})$ is a vector containing all nodal voltages when the domain is excited by $\mathbf{v}_{in}^n$. The error term $(\mathbf{v}^n(\mathbf{p}) - \mathbf{v}_{out}^n)$ represents the mismatch between the simulated ($\mathbf{v}^n(\mathbf{p})$) and desired nodal voltages ($\mathbf{v}_{out}^n$). The superscript $H$ implies the conjugate transpose and the matrix $\bar{\bar{G}}$ is a masking filter used to select and scale elements of the error term. In this work, $\bar{\bar{G}}$ is used to select and scale the nodal voltage mismatch at the output plane, labeled in Fig. 4. Summing (22) over $n$ represents the performance of the device over all input-output pairs, written as,

$$\mathbf{g}(\mathbf{p}) = \sum_{n=1}^{N} g_n(\mathbf{p}). \quad (23)$$

Note that $\mathbf{v}_{out}^n$ fixes the desired nodal voltage magnitudes and phases at selected circuit nodes. Fixing the phases in an absolute sense reduces the solution space and may even present an impossible objective. This is addressed here by updating the fixed phase offset every $m$ iterations according to the simulated output plane nodal voltages, as per the flowchart in Fig. 5. The variable $m$ is selected heuristically and kept fixed through the inverse design procedure. This approach guides the optimization to solutions exhibiting natural phase progression of the fields, typically associated with true time delay (TTD) behavior and gradient-index devices. An alternate objective function may be formulated for beamshaping applications that assigns a desired relative output phase profile, leaving the phase offset arbitrary. This would extend the solution space for TTD designs while avoiding iteration-dependent optimization. This is a topic for future investigations.

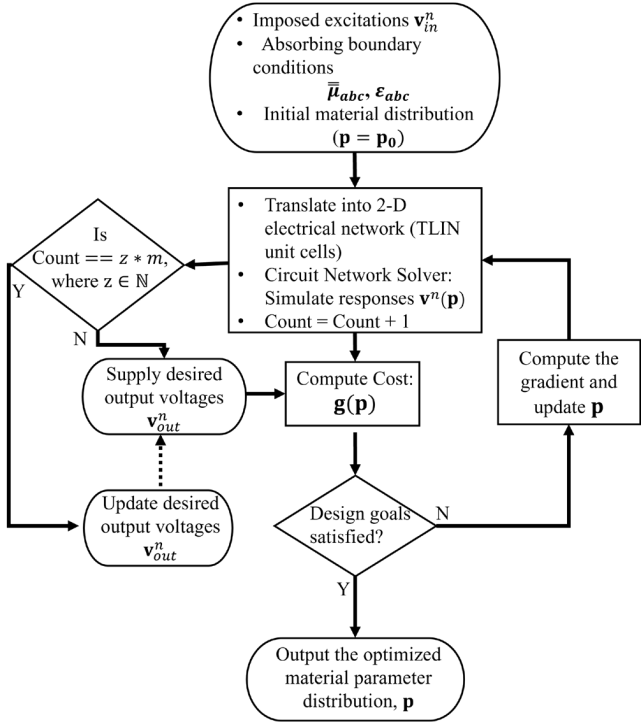

Fig. 5. Flow chart explaining the inverse-design method.

The proposed PMMs offer high-dimensional design spaces where gradient-based optimization methods perform well. However, calculating the gradient of the cost function can be computationally expensive for such design spaces. Therefore, the gradient is calculated using the adjoint-variable method [11], [41], [42]. It is then supplied to the constrained optimization routine fmincon in MATLAB, which updates the design variables of the PMM.

## IV. Design Example I: A Collimator with Amplitude Control

In this section, an inverse-designed PMM that transforms a line source excitation into a collimated wavefront with a cosine amplitude taper is presented. The advantage over conventional TTD beamformers such as Luneburg [46] or Rotman lenses [47] is that the metamaterial can provide precise and independent phase and amplitude control, instead of simply phase control. Additionally, the PMM is appreciably more compact than these earlier multiwavelength-sized devices and supports broadband operation The design methods reported in [28], [37] are able to provide amplitude control, unlike conventional gradient-index devices. However, they rely on intuitive and adiabatic designs, which limit their capability. The inverse-designed collimating PMM reported here is an improvement over the one presented in [27]. In [27], the design possessed a wide range of material parameters ranging from high to near-zero permittivities and permeabilities. The improvements here include using practical material parameters [25] and the introduction of the ABCs in the CNS, which reduces reflection artifacts from the boundaries. It should be noted that the optimized PMM is initially in the form of an electrical network. Then, the device is translated into a lattice of homogeneous blocks with the associated material parameter distribution ($\varepsilon_z(x, y)$ and $\bar{\bar{\mu}}(x, y)$) using (9) and (14). The output voltage and fields extracted from the CNS and COMSOL, respectively, are shown to be in good agreement with the desired field profiles.

### A. Design Specifications

The lossless and reciprocal metamaterial is designed to collimate a line source excitation at 10 GHz. Since the PMM is embedded within free space, the following constraints are imposed,

$$\Delta = 1, \quad \kappa = 1. \tag{24}$$

These two constraints ensure that the PMM is well-matched to the surrounding free space for all illuminations. The PMM is uniformly discretized at a unit cell periodicity $d$, and consists of 31 rows × 20 columns. The unit cell dimension is significantly subwavelength with $d = \lambda_o/12$ in order to reduce spatial dispersion, where $\lambda_o$ is the free space wavelength. The device has a depth of $1.67\lambda_o$ ($20d$ along $\hat{x}$) between the input and output plane, and an output plane/aperture length of $2.6\lambda_o$ ($31d$ along $\hat{y}$). To avoid extreme parameter values, the excitations are displaced from the input plane by $10d$. The desired nodal voltages along the output aperture have a cosine amplitude taper with a flat phase profile at 10 GHz, as plotted in Fig. 6(a). The associated far-field radiation pattern along the azimuthal plane is depicted in Fig. 6(b).

It is evident from (9) and (24) that two free variables ($A$ and $B$) define the electromagnetic properties of each unit cell. They can be related to the local control of power flow and phase progression [28], [36]. Considering all the unit cells in this example, 1240 degrees of freedom are available. The symmetrical nature of the field transformation allows us to condense the design space to 640 unknowns by ensuring the top half mirrors the bottom half of the device. The initial material parameter values of the PMM are set to those of free space. Further, the cost function described in (23) is imposed to penalize a mismatch between the achieved and desired voltage profile at the output plane.

### B. Results

The optimization converged after 200 iterations, and the resulting design and its performance are shown in Fig. 6 and 7. The permeability tensor $\bar{\bar{\mu}}$ in (9) can be recast in terms of a variable $C$, and tensor rotation angle $\Psi$ as follows (Appendix I),

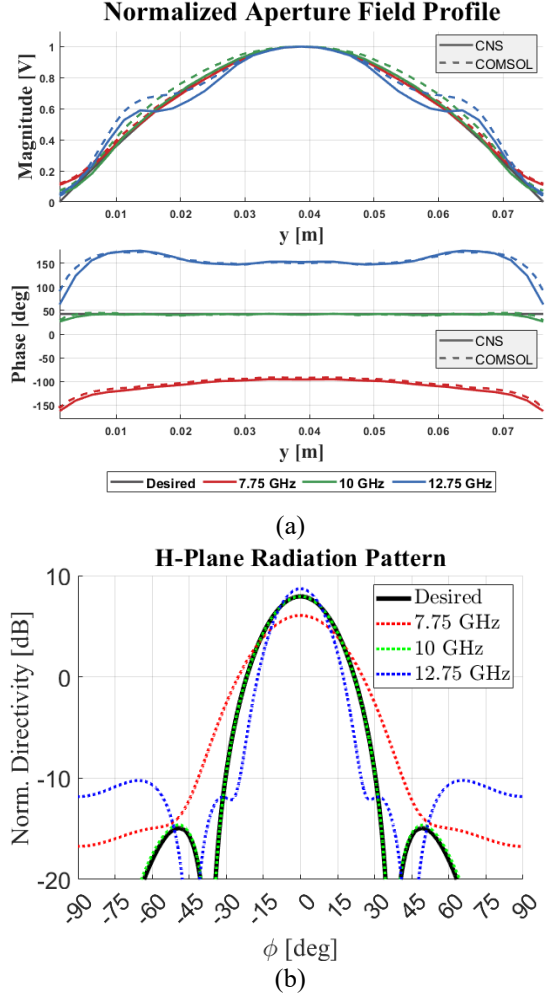

**Fig. 6. Design Example I:** Desired and simulation results of the (a) magnitude and phase profiles of normalized voltage (from CNS) and electric field (from COMSOL) at the output plane of the transformation region (see Fig. 4). (b) Radiation pattern of the PMM beamformer in the azimuthal plane.

$$\bar{\bar{\mu}} = R^T(\Psi)\sqrt{\Delta}\begin{bmatrix} C & 0 \\ 0 & \frac{1}{C} \end{bmatrix} R(\Psi), \quad (25a)$$

$$R(\Psi) = \begin{bmatrix} \cos(\Psi) & \sin(\Psi) \\ -\sin(\Psi) & \cos(\Psi) \end{bmatrix}. \quad (25b)$$

The optimized material parameter distributions are shown in Fig. 7(a)-(d) in terms of $C$, $\Psi$, and $\varepsilon$. The spatial distribution of parameters is clearly inhomogeneous and anisotropic. Various high- and low-index regions are visible that serve to slow down or speed up wave propagation. Fig. 7(e)-(f) plot the complex nodal voltages magnitude and phases within the PMM (represented by the black grid) simulated using the CNS. The source excites a cylindrical wave, composed of a spectrum of plane waves, that is incident on the PMM. The incident wave is focused into several paths. They subsequently recombine to produce a collimated beam with a cosine taper at the aperture. Further, Fig. 7(g) plots the power flow across each unit cell in the top half-plane (mirrored in the bottom half-plane), showing the PMM's refractive control of propagating fields. The PMM collimates 88% of the power entering the input plane along the output plane, 41% more than an equivalently sized free-space region. Additionally, the output fields, computed using the circuit network solver (solid) and COMSOL Multiphysics (dashed), show close agreement with the desired aperture profile, as shown in Fig. 6(a). Consequently, the TLIN implementation of the computational domain in the circuit network solver is verified. Note that COMSOL simulates a 2-D material parameter distribution ($\varepsilon_z(x,y)$ and $\bar{\bar{\mu}}(x,y)$), with no variation assumed along $z$, that is excited by a line source producing S-polarized TEM fields (see Fig. 4(a)). Fig. 6(b) demonstrates close agreement between the desired radiation pattern and the optimized PMM's at the design frequency of 10 GHz. The radiation patterns are calculated by approximating the output aperture as a uniformly spaced array excited by associated complex nodal voltages.

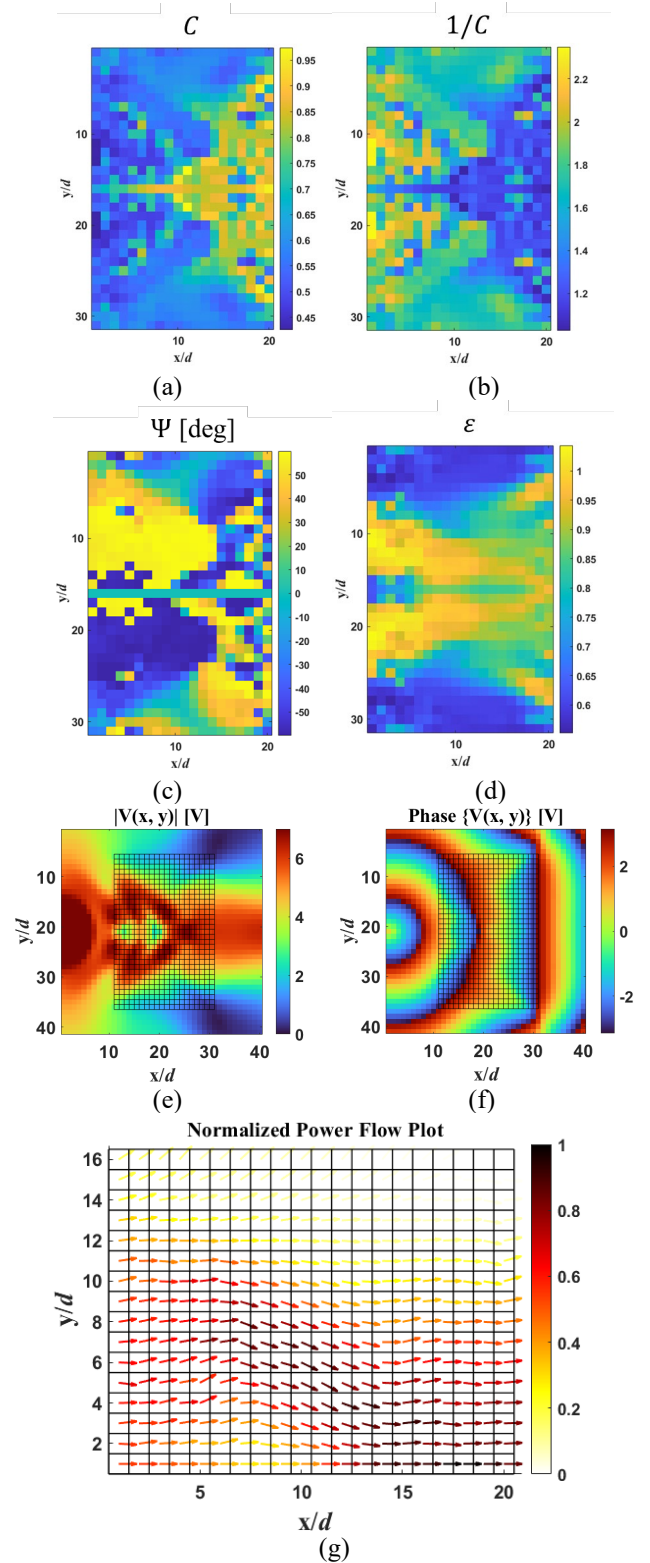

**Fig. 7. Design Example I:** Inverse-designed beam-collimator. (a)-(d) Material parameter distributions. Simulation results of the PMM (represented by the black grid) via CNS for (e) complex nodal voltage magnitudes and (f) complex nodal phases. (g) Power flow within the PMM's top half-plane obtained from the CNS (the colormap represents the magnitude).

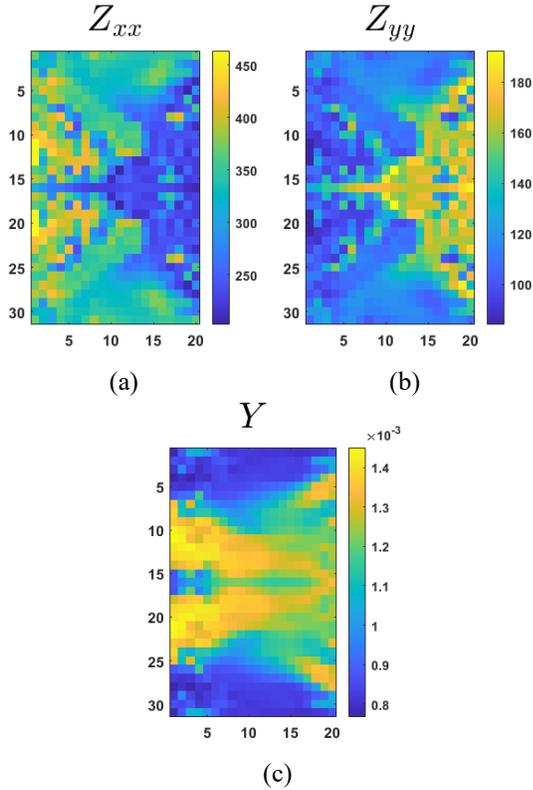

**Fig. 8.** Spatial distribution of TLIN unit cell impedances: (a)-(b) diagonalized impedance tensor entries, (b) shunt susceptance.

A prominent characteristic of TTD devices is broadband operation. A broadband device is typically non-dispersive with a low-pass topology characterized by series inductances and shunt capacitances. The series impedance tensor in (11) can be diagonalized through the following rotation

$$\bar{\bar{Z}} = R^T(\theta) \begin{bmatrix} Z_{xx} & 0 \\ 0 & Z_{yy} \end{bmatrix} R(\theta), \quad (26)$$

The entries of $\bar{\bar{Z}}$ in (26) are plotted in Fig. 8. It can be observed that the series reactances are positive (inductive) and the shunt susceptances are also positive (capacitive). Hence, the device is a low-pass network. The PMM exhibits a 50% bandwidth of operation (7.75 GHz to 12.75 GHz) as observed in Fig. 6(a)-(b).

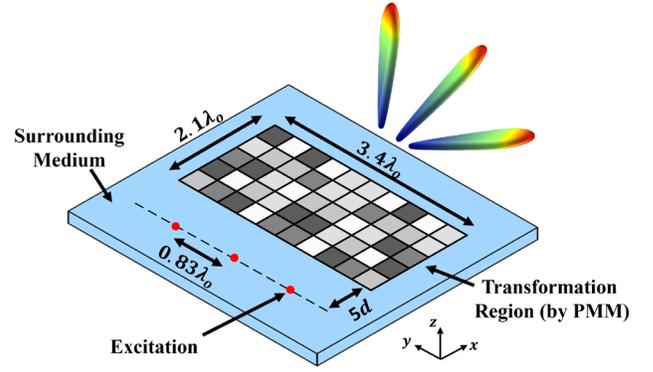

**Fig. 9.** Design example II: an inverse designed 3-Input 3-Output beamformer with zero scan loss.

The PMM's bandwidth is determined by ensuring that sidelobe levels (SLL) are within 3dB of the desired pattern ($SLL_{des} = 22$ dB) and the PMM's efficiency is within 5% compared to the desired frequency ($\eta_{des} = 88\%$). Additionally, the maximum directivity of the main lobe and its half-power beamwidth are maintained within ±1.5 dB and ±20% respectively.

V. DESIGN EXAMPLE: ZERO SCAN LOSS BEAMFORMER

This section describes a 3-input 3-output, PMM beamformer that can azimuthally-scan a cosine-tapered beam with identical directivities, as depicted in Fig. 9. Typically, a loss in directivity is observed as a beam is scanned away from broadside due to a flat aperture's reduction in effective area. The PMM compensates for this scan loss by increasing the aperture size as the beam is scanned away from broadside. Such a design demonstrates that a PMM can produce different aperture field profiles for each input based on TTD operation. This is not possible in conventional analog beamformers such as those based on the Rotman lens [47] or Butler matrix [48]. Butler matrices can realize MIMO systems, however, they are constrained to uniform aperture fields and require multiple components such as phase shifters and hybrid couplers [49]. In contrast, PMMs enable beamforming based on TTD operation with more design flexibility and control over EM fields.

*A. Design Specifications*

The PMM is designed to transform three different point excitations at 10 GHz into beams pointing at the following azimuthal angles: $\theta \in \{-36°, 0°, 36°\}$. The device is embedded within free space, and therefore (24) is imposed. The PMM is a 2-D lattice with periodicity $d = \lambda_o/24$, with 81 rows and 50 columns. The device's depth is $2.1\lambda_o$ (50 cells) and the output plane/aperture is $3.375\lambda_o$ (81 cells). The excitations are placed $5d$ away from the input plane and are spaced $0.83\lambda_o$ apart, in a symmetrical fashion. The desired radiation pattern of the zero scan loss beamformer is plotted in Fig. 10(c). The corresponding aperture field profiles that produce these beams are plotted in Fig. 10(a)-(b) for the $\theta = 0°$ and $-36°$ cases. The fields differ in terms of the aperture width and scaling to ensure a common directivity and radiated power. Note that the desired aperture field for $\theta = 36°$ case set to be symmetrical about the center of the output plane. This allows us to enforce symmetry between the top and bottom half-planes. The resultant design space consists of 4100 variables. The plots for the $\theta = 36°$ case have been excluded since they can be produced by simply mirroring the $\theta = -36°$ plots in Fig. 10(a), 11(e), and 11(f). The initial material parameter values of the PMM are set to those of free space. The cost function, defined in (23), is imposed to penalize a mismatch between the achieved and desired voltage profile at the output plane.

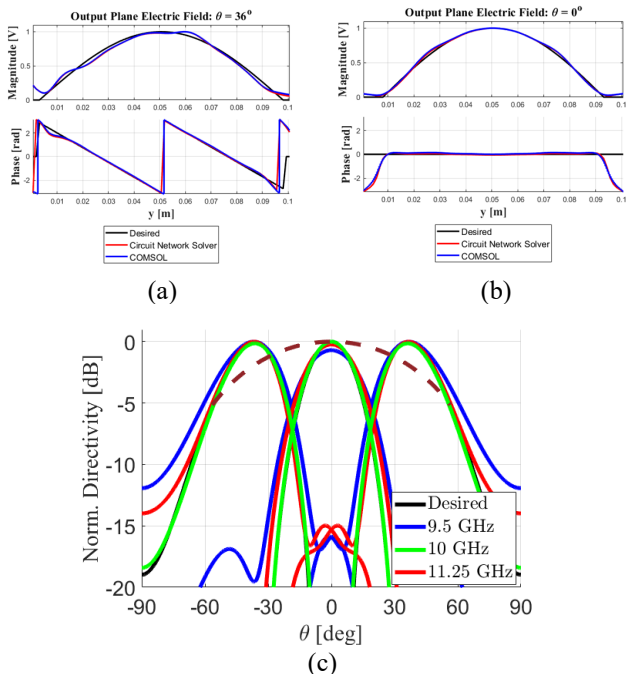

**Fig. 10. Design Example II:** Desired and simulation results of magnitude and phase profiles of normalized voltage (from CNS) and electric field (from COMSOL) at the output plane (see Fig. 4) for (a) $\theta = -36°$, and (b) $\theta = 0°$. (c) Radiation pattern of the PMM beamformer in the azimuthal plane.

*B. Results*

The optimization of the beamformer converged to a solution after 60 iterations. The resulting design and its performance are shown in Fig. 10 and 11. The optimized material parameters are plotted in Fig. 11(a)-(d) and indicate an anisotropic, non-intuitive and non-adiabatic distribution. The beamformer's response (represented by the black grid) for the $\theta = -36°$ case is computed using the CNS and the resultant nodal voltage magnitudes and phases are plotted in Fig. 11(e)-(f). Similarly, Fig. 11(g)-(h) plot the nodal voltages computed for the $\theta = 0°$ case. It is observed that the PMM focuses the incident wave into several paths. This includes focusing fields with large tangential vectors ($k_y$). The propagating fields are subsequently steered and combined to achieve the desired cosine taper and phase profile. The meandering fields within the inverse-designed, highly inhomogeneous and anisotropic device are reminiscent of TTD operation. Evidently, the broadside beam underutilizes the available aperture length. This is key to maintaining the same directivity since the effective aperture area must be increased as the beam is scanned. Such performance is evidenced by the normalized aperture field profiles plotted for both beams in Fig. 10(a)-(b) respectively. As shown, there is close agreement between the desired aperture field profiles and the fields extracted using the CNS as well as COMSOL. Fig. 10(c) plots the radiated patterns at 9.5 GHz, 10 GHz, and 11.25 GHz obtained from the CNS, and the desired radiation pattern against the expected scan loss (dashed brown). The beams are normalized with respect to the maximum directivity at each frequency. The scan loss in each case is reported in Table I. At the design frequency of 10 GHz, the radiation pattern agrees with the desired pattern. The scan loss is 0.15 dB, indicating that it is nearly compensated for when compared to the theoretical scan loss of 1.9 dB. An analysis of the PMM as an electrical network again indicates a low-pass topology throughout. The beamformer exhibits a 1.75 GHz bandwidth (9.5 GHz to 11.25 GHz) as evidenced in Fig. 10(c) and Table I. The bandwidth of operation exhibits a scan loss within 0.5 dB and sidelobe levels better than 15 dB. The performance can be further improved via avenues such as optimizing at multiple frequencies and employing a spectral-based cost function [18]. Reference [25] outlines a roadmap to realizing such broadband PMMs using low-dispersive all-metallic metamaterials.

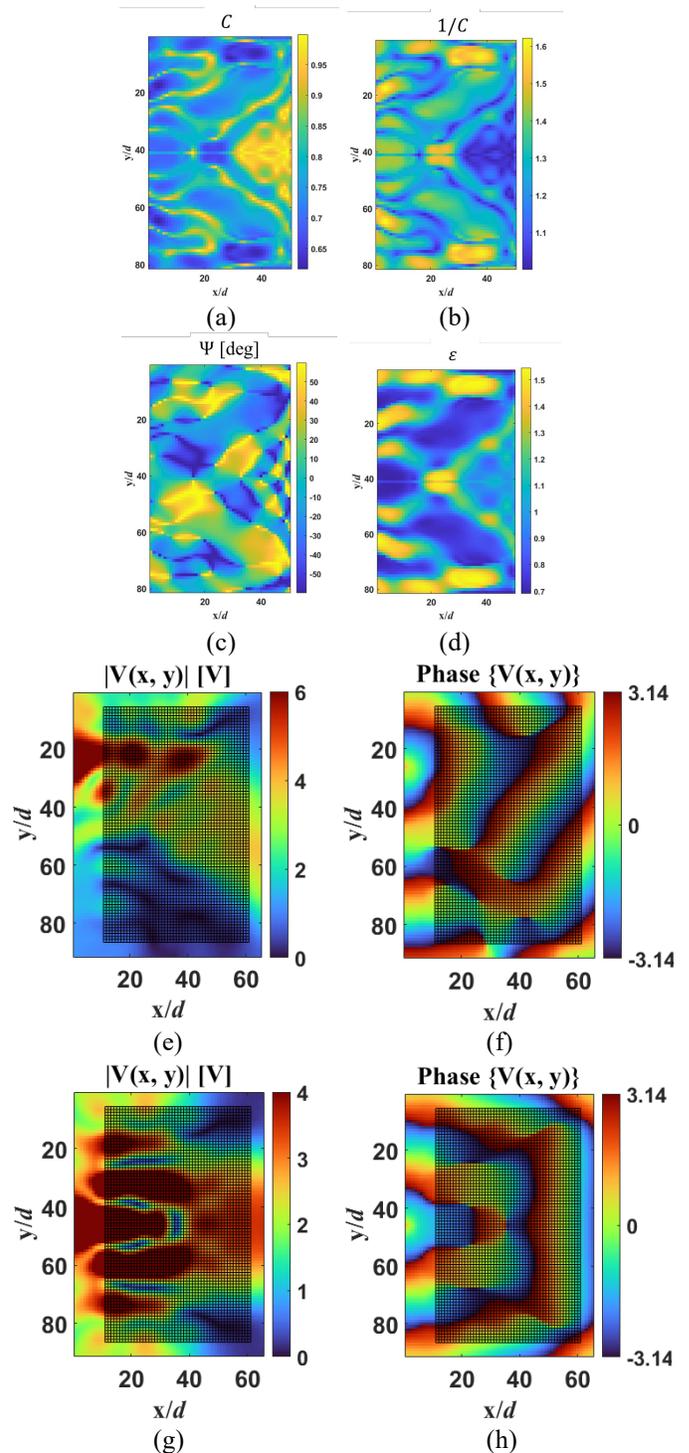

**Fig. 11.** Optimized parameters and results for design example II. (a)-(d) Material parameter distributions. Simulation of the PMM (represented by the black grid) via CNS for (e) complex nodal voltage magnitudes and (f) complex nodal phases for $\theta = -36°$. (g) Complex nodal voltage magnitudes and (h) complex nodal phases for $\theta = 0°$.

TABLE I
SCAN LOSS ACROSS BANDWIDTH

| Frequency (GHz) | Scan Loss (dB) |
|---|---|
| Theoretical (20 log(cos 36°)) | 1.9 |
| 9.5 | 0.5 |
| 10 | 0.15 |
| 11.25 | 0.3 |

## VI. CONCLUSION

In this paper, an inverse-design procedure for designing multi-input multi-output (MIMO) 2-D perfectly-matched metamaterials (PMMs) was presented. PMMs are discretized, inhomogeneous, and anisotropic metamaterials that mitigate internal reflections and are impedance-matched with the surrounding medium. The PMMs exert refractive control over propagating fields and promise broadband performance in low-dispersive media. A transmission-line (TLIN) metamaterial unit cell that exhibits magnetic anisotropy was introduced to model PMM unit cells. PMMs excited by S-polarized TEM plane waves are represented by 2-D electrical networks composed of these TLIN unit cells. The proposed inverse design method employs a circuit-based forward solver to rapidly compute the response of the device within a computational domain lined with TLIN-based absorbing boundary conditions. The numerical accuracy of the solver is verified through comparisons with a commercial full-wave solver. The circuit network solver is coupled with a gradient-based optimization routine that is accelerated by the adjoint variable method.

The advantages offered by inverse-designed PMMs were demonstrated through the design of beamforming transformation regions such as a planar beam-collimator and MIMO beamformer at 10 GHz. The beam collimator transforms the incident field of a line source excitation into an output field with a cosine amplitude taper and collimated phase with high efficiency and across a 50% bandwidth from 7.75 GHz to 12.75 GHz. The MIMO beamformer supports the simultaneous excitation of three beams with equal directivities having a 1.75 GHz bandwidth from 9.5 GHz to 11.25 GHz. It compensates for the scan loss by varying the effective aperture size for each beam while maintaining the same radiated power.

The proposed inverse-design method is suitable for wideband and multifunctional metamaterial designs for applications such as broadband signal routing, lensing, beamforming, and analog signal processing. Future work includes 1) investigating objective functions that allow free-form aperture fields (normalized magnitudes and arbitrary DC phase-offsets) that extend the TTD solution space, 2) multi-frequency inverse-design, and 3) implementation of broadband PMMs using low-dispersion, all-metal metamaterials.

## APPENDIX A

ALTERNATE FORMULATION OF PERFECTLY-MATCHED MEDIA

This appendix relates the two formulations for the material parameters of perfectly-matched media provided in (9) and (25).

The expression in (25a) can be simplified to

$$\bar{\bar{\mu}} = \sqrt{\Delta} \begin{bmatrix} C\cos^2(\Psi) + \frac{1}{C}\sin^2(\Psi) & \sin(\Psi)\cos(\Psi)\left(C - \frac{1}{C}\right) \\ \sin(\Psi)\cos(\Psi)\left(C - \frac{1}{C}\right) & C\sin^2(\Psi) + \frac{1}{C}\cos^2(\Psi) \end{bmatrix}. \quad (27)$$

The permittivity can be derived using (8) as the following

$$\varepsilon_z = \frac{\kappa}{\sqrt{\Delta}\left(C\cos^2(\Psi) + \frac{1}{C}\sin^2(\Psi)\right)}. \quad (28)$$

Comparing the permeability tensor in (9) with (26), we arrive at the following relationship between the unit cell variable pairs $(A, B)$ and $(C, \Psi)$:

$$A = \frac{1}{\sqrt{\Delta}\left(C\cos^2(\Psi) + \frac{1}{C}\sin^2(\Psi)\right)}, \quad (29a)$$

$$B = \sqrt{\Delta}\left(\sin(\Psi)\cos(\Psi)\left(C - \frac{1}{C}\right)\right). \quad (29b)$$